\documentclass[conference]{IEEEtran}

\usepackage{amsmath,amssymb,bm}
\usepackage{array}
\usepackage{algorithm}
\usepackage{algorithmic}
\usepackage{booktabs}
\usepackage{cite}
\usepackage{graphicx}
\usepackage[acronym]{glossaries}
\usepackage[table]{xcolor}

\newcommand{\E}{\mathbb{E}}

\newcommand{\tx}{\mathrm{tx}}

\newcolumntype{C}[1]{>{\centering\arraybackslash}p{#1}}

\newacronym{isac}{ISAC}{Integrated Sensing and Communication}
\newacronym{an}{AN}{Artificial Noise}
\newacronym{awgn}{AWGN}{Additive White Gaussian Noise}
\newacronym{ber}{BER}{Bit Error Rate}
\newacronym{ldpc}{LDPC}{Low-Density Parity-Check}
\newacronym{qpsk}{QPSK}{Quadrature Phase-Shift Keying}
\newacronym{ofdm}{OFDM}{Orthogonal Frequency-Division Multiplexing}
\newacronym{rf}{RF}{Radio Frequency}
\newacronym{5gnr}{5G NR}{Fifth-generation New Radio}
\newacronym{uav}{UAV}{Unmanned Aerial Vehicle}
\newacronym{los}{LOS}{Line-of-Sight}
\newacronym{mcs}{MCS}{Modulation and Coding Scheme}
\newacronym{cu}{CU}{Communication User}
\newacronym{bs}{BS}{Base Station}
\newacronym{snr}{SNR}{Signal-to-Noise Ratio}
\newacronym{psr}{PSR}{Peak Sidelobe Ratio}
\newacronym{crb}{CRB}{Cramer-Rao Bound}
\newacronym{glrt}{GLRT}{Generalized Likelihood Ratio Test}
\newacronym{ser}{SER}{Symbol Error Rate}
\title{Waveform Randomization for Secure ISAC}

\author{
\IEEEauthorblockN{Zexin~Fang\IEEEauthorrefmark{1},~Bin~Han\IEEEauthorrefmark{1},
Shuangyang~Li\IEEEauthorrefmark{3},
and~Hans~D.~Schotten\IEEEauthorrefmark{1}\IEEEauthorrefmark{2}}
\IEEEauthorblockA{\IEEEauthorrefmark{1}{RPTU University Kaiserslautern-Landau} and 
\IEEEauthorrefmark{2}{German Research Center for Artificial Intelligence (DFKI), Germany}\\
\IEEEauthorrefmark{3}{Technische Universit\"at Berlin, Germany}}
}

\begin{document}
\maketitle

\begin{abstract}
This paper studies waveform-level security for \gls{isac}. Instead of
relying on spatial beamforming or power allocation, we randomize the
sensing waveform itself using phase keys and tangent \gls{an} applied to a
Fourier-curve constellation. \textcolor{black}{Coordinated legitimate nodes share the per-burst
phase key and the \gls{an} realization and use them to reconstruct the sensing
reference, whereas other receivers observe a keyed waveform whose \gls{an}
realization is unknown to them.} We
consider a \gls{uav}-mounted Eve with a dominant \gls{los} link who fully
exploits favorable geometry, including a bypass scenario where Eve listens
to both the direct and echo sides to circumvent our design. \textcolor{black}{We further
consider a stronger Eve that attempts phase-key reconstruction from
long-term observations; even then, residual key mismatch structurally
degrades correlation and \gls{an} cancellation.} Numerical results show that the
proposed design reduces Eve's probability of detecting the sensing
reference using non-coherent detection and, in the bypass case,
substantially suppresses her correct-delay lock probability, while the
legitimate receiver reliably locks onto the reference through echo
accumulation.
\end{abstract}
\glsresetall

\begin{IEEEkeywords}
\glsentryshort{an}, \glsentryshort{isac}, B5G, 6G, Security. 
\end{IEEEkeywords}

\section{Introduction}

\gls{isac} is a promising paradigm for next-generation wireless networks,
reusing radio resources for both data transmission and environmental
sensing within a unified system~\cite{liu2020isac}, thereby improving
spectral efficiency and reducing system cost toward large-scale,
sustainable sensing services. However, its dual-purpose design introduces a physical-layer security
risk: parties aware of the resource structure, whether external
eavesdroppers or authorized \glspl{cu}, may monitor sensing
activity or exploit repeated sensing references to perform unauthorized
tracking or identification. 

\gls{an} is a classical way to degrade an unintended receiver without
relying solely on higher-layer
encryption~\cite{goel2008artificial}. Most secure-\gls{isac} designs
combine \gls{an} with beamforming, steering useful energy toward legitimate
nodes while injecting \gls{an} into the remaining spatial directions to
degrade
Eve~\cite{zou2024securing,xu2022robust,su2024sensing,jia2024physical,ren2023robust},
with some made robust to bounded uncertainty in her location or
channel~\cite{xu2022robust,ren2023robust}, and others jointly optimizing
beamforming and \gls{an} covariance to make multi-target echoes less
distinguishable~\cite{huang2026isacprivacy}. However, a \gls{uav}-borne Eve is not confined to a bounded region:
she can loiter, hover, and actively reposition to explore the
surrounding geometry, converging on a vantage point with a dominant
\gls{los} link and favorable beamforming gain. Ambiguity-function
engineering offers an alternative via ghost
targets~\cite{han2026sensingsecure}, but these decoys neither hide the real
echo nor conceal the sensing activity itself, so prolonged observation still
lets Eve isolate the true target.

This motivates our central question: can the sensing reference be hidden
from Eve while remaining decodable to the legitimate receiver, e.g., a
cooperating \gls{bs}? Unlike spatial-domain defenses that control where
energy is radiated, we protect what waveform is observed. We use a Fourier-curve constellation
protected by phase keys and tangent \gls{an}: the legitimate receiver,
coordinated over secure backhaul (BS--BS), knows the per-burst phase key
and recovers the reflected sensing waveform, while Eve or a compromised
\gls{cu}, holding only the communication-domain key, sees only a
randomized waveform. Meanwhile, we consider an \gls{uav}-borne Eve who,
exploiting a strong \gls{los} link to the transmitter and the target, can
bypass the waveform-domain protection by comparing the direct
and reflected signals, without needing to recover the phase key. 
% \gls{an} realization. This represents a favorable, near-worst-case model
% for Eve: rather than a weak, out-of-beam listener, she is granted a
% dominant propagation advantage and architectural knowledge of the system.
Without the phase keys needed to isolate the true sensing reference from the
transmitted structure, Eve, though able to exploit favorable geometry,
remains limited to single-echo observation: her correct-delay lock
\textcolor{black}{probability stays low, whereas the legitimate receiver coherently integrates
echoes using the shared schedule and known reference.} With phase-key
reconstruction, Eve can perform coherent integration across bursts, but her
\textcolor{black}{performance remains limited by residual phase-key error and imperfect
\gls{an} removal.}

We consider an \gls{ofdm} \gls{isac} system using \gls{qpsk} for
sensing/control symbols and higher-order modulation for payload under link
adaptation, consistent with \gls{5gnr} \cite{3gpp38211}. The rest of this
paper is organized as follows. Section~II presents the secure communication
and sensing model. Section~III describes the threat model and detection
metrics. Section~IV reports the numerical results, and Section~V concludes
the paper. 

\section{Secure Communication and Sensing Model}

\subsection{Fourier-Curve Modulator}
Let $M=2^q$. Every group of $q$ coded bits selects an index
$i\in\{0,\ldots,M-1\}$, mapped to a phase
$\theta_i=\frac{2\pi i}{M}$.

We map the conventional $M$-PSK phase $\theta_i$ to a $k$-dimensional
harmonic symbol vector,
\begin{equation}
\bm{x}_i=\frac{1}{\sqrt{k}}
\begin{bmatrix}
        e^{j\theta_i} & e^{j2\theta_i} & \cdots & e^{jk\theta_i}
\end{bmatrix}^{\mathsf T},
\qquad \|\bm{x}_i\|=1.
\label{eq:curve_symbol_vector}
\end{equation}
The $k$ components are jointly determined by the single index $i$, and may
be mapped to parallel subcarriers, orthogonal signaling dimensions, or
successive resource elements depending on the physical implementation.

Differentiating $\bm{x}_i$ with respect to $\theta_i$ traces the local
direction of the symbol-vector curve as $i$ varies, with
$\bm{x}'(\theta_i)=\frac{j}{\sqrt{k}}
\begin{bmatrix}
        e^{j\theta_i} & 2e^{j2\theta_i} & \cdots &
        ke^{jk\theta_i}
\end{bmatrix}^{\mathsf T}$. Normalizing by its norm $v_k=\sqrt{\frac{1}{k}\sum_{\ell=1}^{k}\ell^2}$, 
gives the unit tangent vector at $\bm{x}_i$,
\begin{equation}
\hat{\bm{t}}_i=\frac{\bm{x}'(\theta_i)}{v_k},
\qquad \|\hat{\bm{t}}_i\|=1.
\label{eq:unit_tangent}
\end{equation}
The tangent is orthogonal to the symbol vector,
$\Re\{\bm{x}_i^{\mathsf H}\hat{\bm{t}}_i\}=0$, once each complex vector is
identified with its real embedding in $\mathbb{R}^{2k}$. This confirms that
\textcolor{black}{$\hat{\bm{t}}_i$ describes a purely rotational (phase) perturbation of
$\bm{x}_i$, with no component along $\bm{x}_i$ itself.}
\subsection{Tangent \glsentryshort{an} and Channel}
Before \gls{an} injection, the transmitter applies a shared phase-key vector
\(
\bm{\varphi}=[\varphi_1,\ldots,\varphi_k]^{\mathsf T}
\)
across the $k$ harmonic components, giving the keyed Fourier-curve symbol vector
\begin{equation}
\bm{x}_i(\bm{\varphi})
    =\frac{1}{\sqrt{k}}
\begin{bmatrix}
        e^{j(\theta_i+\varphi_1)} &
        e^{j(2\theta_i+\varphi_2)} &
\cdots &
        e^{j(k\theta_i+\varphi_k)}
\end{bmatrix}^{\mathsf T}.
\label{eq:keyed_symbol_vector}
\end{equation}
The keyed tangent $\hat{\bm{t}}_i(\bm{\varphi})$ follows by differentiating
$\bm{x}_i(\bm{\varphi})$ with respect to $\theta_i$. The legitimate receiver
knows $\bm{\varphi}$ and builds the keyed codebook directly; an
unintended receiver without the key faces a geometry mismatch even before
\gls{an} is considered.

For transmission $r$, the transmitter draws one real scalar
$\xi_r\sim\mathcal N(0,1)$, shared across all $k$ harmonic components
\begin{equation}
\bm{x}^{\tx}_{i,r}
    =\sqrt{1-\beta}\,\bm{x}_i(\bm{\varphi})
    +\sqrt{\beta}\,\xi_r\hat{\bm{t}}_i(\bm{\varphi}),
\qquad 0\leq\beta<1.
\label{eq:protected_symbol_vector}
\end{equation}
The tangent direction is deterministic for a fixed $i$, while the signed
amplitude $\xi_r$ is random and drawn independently for each
symbol vector, not each component. Since $\bm{x}_i(\bm{\varphi})$ and
$\hat{\bm{t}}_i(\bm{\varphi})$ are unit-norm and orthogonal with
$\E\!\left[\|\bm{x}^{\tx}_{i,r}\|^2\right]=1$.

After synchronization and channel equalization, the real-valued received
symbol vector is
\begin{equation}
\bm{Y}_r=\sqrt{1-\beta}\,\bm{x}_i(\bm{\varphi})
    +\sqrt{\beta}\,\xi_r\hat{\bm{t}}_i(\bm{\varphi})+\bm{N}_r,
\label{eq:received_model}
\end{equation}
where $\bm{Y}_r,\bm{x}_i,\hat{\bm{t}}_i,\bm{N}_r\in\mathbb{R}^{2k}$ and
$\bm{N}_r\sim\mathcal N(\bm{0},\sigma_c^2\bm{I}_{2k})$. The total symbol-vector
\textcolor{black}{SNR is}
$\rho=\frac{\E[\|\bm{x}^{\tx}_{i,r}\|^2]}
    {\E[\|\bm{N}_r\|^2]}=\frac{1}{2k\sigma_c^2}
\label{eq:snr_definition}$.

\subsection{QPSK Sensing-Waveform Design}
We consider QPSK phase sequences
\begin{equation}
s[n] = e^{j\theta_n}, \theta_n \in \left\{0, \frac{\pi}{2}, \pi, \frac{3\pi}{2}\right\}, n = 0, \ldots, N-1.
\label{eq:qpsk}
\end{equation}

% Review: Fig. 4 packs 320 coefficients into one OFDM block; reconcile this mapping and define R_k before the objective.
\textcolor{black}{Each $s[n]$ is mapped to the weighted subcarrier symbol vector
$\bm{u}_k[n]$, as demonstrated in Eq.~\ref{eq:curve_symbol_vector}. Its entries are loaded onto the active subcarriers $\mathcal{K}$ of an $N_{\rm FFT}$-point OFDM symbol (inactive subcarriers set to zero) and converted to the time-domain sensing burst $\bm{x}_n$ by the unitary inverse OFDM transform.\footnote{That is, $\bm{x}_n = \bm{F}^{\mathsf H}\bm{U}_n$, where $\bm{U}_n\in\mathbb{C}^{N_{\rm FFT}}$ satisfies $\bm{U}_n[\mathcal{K}]=\bm{u}_k[n]$, $\bm{U}_n[\mathcal{K}^c]=\bm{0}$, and $\bm{F}$ is the unitary DFT matrix.} Because $\bm{F}$ is unitary and $\bm{U}_n$ is supported only on $\mathcal{K}$, Parseval's theorem gives $\bm{u}_k[n]^{\mathsf H}\bm{u}_k[n-\tau] = \bm{x}_n^{\mathsf H}\bm{x}_{n-\tau}$, so correlating the weighted subcarrier symbol vectors is equivalent to correlating the transmitted time-domain bursts. The resulting subcarrier-combined harmonic autocorrelation
}
\begin{equation}
J_k(\bm{\theta})
=
\sum_{\tau\neq 0}|R_k[\tau]|^2
+
\lambda
\max_{\tau\neq 0}|R_k[\tau]|^2 ,
\label{eq:qpsk_objective}
\end{equation}
where $\bm{\theta}=[\theta_0,\ldots,\theta_{N-1}]^{\mathsf T}$ collects the \gls{qpsk} phases of the sensing sequence. The first term penalizes total sidelobe energy and the second the peak sidelobe, with $\lambda$ controlling their tradeoff; we set $\lambda=10$ throughout to emphasize \gls{psr}. Since the phase alphabet is discrete, the resulting optimization is combinatorial; we solve it via discrete coordinate descent with random restarts, repeated per harmonic order $k$ and summarized in Algorithm~\ref{alg:qpsk_waveform_design}.

\textcolor{black}{The nominal sensing sequence is phase-keyed and protected by tangent
\gls{an}. Cooperating BSs synchronize and share the burst schedule, sensing
key $\bm{\varphi}_{\mathrm{s}}$, and \gls{an} realization $\xi$ over backhaul.
The receiver correlates the known reference $\bm{x}$ over the
expected delay window.} After compensating known front-end and
channel distortions, the sensing waveform is reconstructed, with remaining
uncertainty from residual timing error, further reduced through
echo accumulation. For communication, the \gls{bs} and \gls{cu} instead share a
communication-domain key $\bm{\varphi}_{\mathrm{c}}$ \cite{han2026matched}, instantiating the
same keyed lookup table of Fourier-curve symbol vectors and tangent directions,
enabling covariance-aware demodulation without knowledge of the
\gls{an} realization. This is to protect the sensing
reference from both external eavesdroppers and \glspl{cu}.

\begin{algorithm}[b]
\scriptsize
\caption{\glsentryshort{qpsk} harmonic-waveform design for fixed $k$}
\label{alg:qpsk_waveform_design}
\begin{algorithmic}[1]
\STATE Choose $N$, $k$, restarts $R$, sweeps $S$; define $\mathcal{Q}=\{0,\pi/2,\pi,3\pi/2\}$ and objective $J_k(\bm{\phi})$ in \eqref{eq:qpsk_objective}.
\FOR{$r=1,\ldots,R$}
\STATE Initialize $\bm{\phi}^{(r)}$ (structured or random seed).
\FOR{$s=1,\ldots,S$}
\STATE \texttt{improved} $\leftarrow$ false.
\FOR{each chip $n$ (random order)}
\STATE Set $\phi_n\leftarrow\arg\min_{\phi\in\mathcal{Q}}J_k$, others fixed; if $J_k$ decreased, \texttt{improved} $\leftarrow$ true.
\ENDFOR
\IF{\texttt{improved} = false}
\STATE \textbf{break} (stop restart $r$).
\ENDIF
\ENDFOR
\STATE Store best sequence from restart $r$.
\ENDFOR
\STATE Return the best sequence overall (or all within tolerance of it).
\end{algorithmic}
\end{algorithm}
\subsection{Covariance-Aware Demodulation}

\subsubsection{Unknown-\glsentryshort{an} Maximum Likelihood Receiver}
Unlike the sensing case, where the waveform $\bm{x}$ is known, communication has no such prior: the candidate
$i$ is itself the unknown, so its tangent direction $\hat{\bm{t}}_i$ is
unknown until $i$ is decided. The receiver therefore marginalizes $\xi_r$
rather than estimating it, giving
\begin{equation}
\bm{Y}_r\mid i\sim\mathcal N\!\left(
\sqrt{1-\beta}\bm{x}_i,\bm{\Sigma}_i
\right),
\end{equation}
with $\bm{\Sigma}_i=\sigma_c^2\bm{I}_{2k}
    +\beta\hat{\bm{t}}_i\hat{\bm{t}}_i^{\mathsf T}
\label{eq:covariance}$.

The \gls{an} increases variance only along candidate $i$'s tangent. Since all
tangents have unit norm, with
$\det(\bm{\Sigma}_i)
    =(\sigma_c^2)^{2k-1}(\sigma_c^2+\beta)$
is independent of $i$. Define $\bm{r}_i=\bm{Y}_r-\sqrt{1-\beta}\bm{x}_i$.
Applying the Sherman--Morrison identity yields the candidate metric
\begin{equation}
\Lambda_i(\bm{Y}_r)
    =\frac{\|\bm{r}_i\|^2}{\sigma_c^2}
    -\frac{\beta(\bm{r}_i^{\mathsf T}\hat{\bm{t}}_i)^2}
    {\sigma_c^2(\beta+\sigma_c^2)}.
\label{eq:ml_metric}
\end{equation}
The hard decision is $\hat{i}=\arg\min_i\Lambda_i$. The first term is the
ordinary Euclidean metric; the second reduces the penalty when the residual
lies along the \gls{an} direction expected under candidate $i$. The receiver thus
accounts for the candidate-dependent covariance rather than erasing or
estimating the \gls{an} sample, a correction the Euclidean baseline omits.

Let $b_a(i)$ denote bit position $a\in\{0,\ldots,q-1\}$ in the binary label
of candidate $i$. For coded control and payload channels, the
covariance-aware soft demapper forms
\begin{equation}
L_{r,a}=\log
\frac{\sum_{i:b_a(i)=0}\exp[-\Lambda_i(\bm{Y}_r)/2]}
{\sum_{i:b_a(i)=1}\exp[-\Lambda_i(\bm{Y}_r)/2]} .
\label{eq:llr}
\end{equation}
Both sums are evaluated with log-sum-exp for numerical stability; positive
$L_{r,a}$ favors bit zero. \textcolor{black}{A rate-$1/2$ \gls{ldpc} code with}
belief-propagation decoding is used only for control and payload, which
have explicit \gls{ber} targets. \textcolor{black}{The sensing waveform remains uncoded to
preserve the optimized phase sequence; the receiver instead correlates with
the known sensing reference.}

\section{Threat Model}\label{sec:thmodel}
We assume Eve (i) performs non-coherent detection on the direct and
reflected signals to bypass the system, (ii) exploits long-term
observations to attempt cross-burst coherent reference reconstruction,
\textcolor{black}{and (iii) reconstruct phase-keys.} If the
key estimate is sufficiently accurate, Eve can recover the nominal
sensing sequence and form a coherent reference. This is a stronger
assumption than Eve merely knowing the waveform-generation structure,
since a newly generated nominal-like sequence remains misaligned with the
observed protected waveform.
\textcolor{black}{Cyclostationary detection based on repeated burst
timing can be mitigated by randomizing the sensing-burst schedule, which is
shared with the legitimate receiver over secure backhaul.} Since \gls{ofdm} subcarrier spacing is public
numerology observable from the direct link, Eve could apply standard
Doppler compensation without knowledge of the sensing sequence; moreover,
for typical target/\gls{uav} velocities the Doppler shift is small
relative to the subcarrier spacing regardless. We therefore adopt a
delay-only detection model.

\gls{uav}-to-\gls{uav} and
\gls{uav}-to-ground links at typical operating altitudes and
frequencies are predominantly \gls{los}, with multipath components
attenuated well below the direct path due to the elevated,
obstruction-free geometry. Under this assumption, both the direct and
\textcolor{black}{reflected links are well approximated as single-tap
(\glsentryshort{los}-dominant) channels. This LOS-only, single-tap model
represents a best-case channel for Eve and thus yields a favored detection performance for Eve.} With
\begin{equation}
h_d[n]\approx g_d\,\delta[n-\tau_d],
\quad
h_r[n]\approx g_r\,\delta[n-\tau_r],
\end{equation}
where $g_d,g_r$ are the complex path gains and $\tau_d,\tau_r$ are fixed
propagation delays. Absorbing $\tau_d$ into the time reference and defining
$\ell_0\triangleq\tau_r-\tau_d$ as the delay of interest, the convolutions
$(h_d*x)[n]$ and $(h_r*x)[n-\ell_0]$ collapse to simple scaled shifts,
$g_d\,x[n]$ and $g_r\,x[n-\ell_0]$, respectively. Under a delay-only bypass
model, the direct-path signal and reflected-path signal are $d[n]$ and $r[n]$, and the two hypotheses are then
\begin{equation}\begin{split}
\mathcal{H}_0:&
\begin{cases}
d[n]=g_d\,x[n]+u[n],\\
r[n]=w[n],
\end{cases}\\
\mathcal{H}_1:
&\begin{cases}
d[n]=g_d\,x[n]+u[n],\\
r[n]=g_r\,x[n-\ell_0]+w[n],
\end{cases}\end{split}
\label{eq:eve_bypass_hyp}
\end{equation}
where $\mathcal{H}_0$ is the null hypothesis that no target-generated
echo is present at Eve, $\mathcal{H}_1$ is the target-present
hypothesis, $x[n]$ is the transmitted sensing burst, and
$u[n],w[n]\sim\mathcal{CN}(0,\sigma_w^2)$ are independent \gls{awgn} terms.
Under $\mathcal{H}_0$, the detection threshold is set for the required
false-alarm rate $P_{\mathrm{FA}}$. We then assess under $\mathcal{H}_1$ whether
Eve detects the target echo and estimates its delay.
%This LOS-dominant simplification is consistent with the delay-only detection
%model adopted throughout this subsection, since a frequency-selective $h_d$
%or $h_r$ would introduce additional delay spread that conflicts with treating
%$\ell_0$ as the single unknown delay of interest.

Eve uses the direct path as a
surrogate and computes the cross-correlation statistic
\begin{equation}
    z[\ell]=\sum_n r[n]\,d^*[n-\ell],
\qquad \ell=0,\dots,L-1,
\label{eq:matched_filter_delay}
\end{equation}
followed by the peak-search rule
\begin{equation}
\max_{\ell} |z[\ell]| \underset{\mathcal{H}_0}{\overset{\mathcal{H}_1}{\gtrless}} \eta,
\qquad
\hat\ell=\arg\max_\ell |z[\ell]|.
\label{eq:peak_search}
\end{equation}
$\eta$ is the non-coherent detection
threshold applied to the correlation magnitude.
\textcolor{black}{Since $\ell_0$ is neither known nor grid-aligned}, Eve must search candidate
bins; sidelobes then affect only the rate of spurious crossings at incorrect
bins, not the correct-bin statistic itself, so the two effects are treated
separately below. Define the normalized autocorrelation of the sensing
burst,
\begin{equation}
\chi[\ell] \triangleq \frac{1}{E_s}\sum_n x[n]\,x^*[n-\ell],
\qquad E_s=\sum_n|x[n]|^2.
\label{eq:ambiguity_def}
\end{equation}

\subsubsection{Noise Model} $z[\ell]$ correlates two noisy signals rather
than one noisy signal against a clean reference.
The noise cross-term $w[n]u^*[n]$ contributes equally under both hypotheses,
so any asymmetry in $\sigma_z^2$ would have to come from the signal-noise
cross term $g_r\sum_n x[n-\ell_0]u^*[n]$ under $H_1$; We further restrict attention
to the regime of a strong direct link, so that
$d[n]\approx g_d x[n]$ acts as a quasi-clean surrogate reference and this
cross-noise term is negligible. Under this approximation the noise variance
is dominated by the echo channel noise variance $\sigma_w$ alone, with
$\sigma_z^2 \approx |g_d|^2\sigma_w^2 E_s$, 
which recovers the standard complex-Gaussian form
\begin{equation}\begin{split}
    &z[\ell_0]\mid \mathcal{H}_0 \sim \mathcal{CN}(0,\sigma_z^2),\\
    &z[\ell_0]\mid \mathcal{H}_1 \sim \mathcal{CN}(g_rg_d^* E_s\chi[0],\,\sigma_z^2).
\end{split}\label{eq:pd_statistic}
\end{equation}

\subsubsection{Correct-Bin and False-Lock Probabilities}
As before,
\begin{equation}\begin{split}
P_{\mathrm{corr}}&=\Pr\!\big(\hat{\ell}=\ell_0,\ |z[\hat{\ell}]|>\eta
\mid\mathcal{H}_1\big),\\
P_{\mathrm{false}}&=\Pr\!\big(\hat{\ell}\neq \ell_0,\ |z[\hat{\ell}]|>\eta
\mid\mathcal{H}_1\big),
\label{eq:pcorr_pfalse}\end{split}
\end{equation}
with $P_D=P_{\mathrm{corr}}+P_{\mathrm{false}}$ and
$P_{\mathrm{miss}}=1-P_D$. 

\subsubsection{Role of the Mainlobe}
Ignoring competition from other bins, the
\textcolor{black}{mainlobe alignment $|\chi[0]|^2$} bounds correct-bin detection,
\begin{equation}
    P_{\mathrm{corr}} \;\lesssim\;
    Q_1\!\left(
\sqrt{\frac{2|g_rg_d|^2 E_s^2\,|\chi[0]|^2}{\sigma_z^2}},
\textcolor{black}{\sqrt{-2\ln P_{\mathrm{FA}}}}
\right),
\label{eq:pd_explicit}
\end{equation}
which reduces to the classical matched-filter result under a perfectly clean
direct link and ideal waveform, and is an upper bound rather than the true $P_{\mathrm{corr}}$.

\subsubsection{Role of the Sidelobes}
$P_{\mathrm{false}}$ is governed by the
off-peak values $\chi[\ell]$, $\ell\neq0$: a large sidelobe at offset
$\ell'\neq0$ gives mean response $g_rg_d^*E_s\chi[\ell']$ at bin
$\ell_0+\ell'$, which can cross $\eta$ and inflate $P_{\mathrm{false}}$, and
hence $P_D$, without improving ranging accuracy. A lower-sidelobe
waveform can exhibit \emph{lower} $P_D$ than a higher-sidelobe one while
achieving equal or higher $P_{\mathrm{corr}}$, and why the peak sidelobe
ratio,
$\mathrm{PSR}=\frac{1}{\max_{\ell\neq0}|\chi[\ell]|^2},
\label{eq:psr_def}$
\textcolor{black}{together with $|\chi[0]|^2$}, characterizes ranging reliability more
completely than $P_D$ alone. 
\begin{figure*}[!t]
\centering
\includegraphics[width=0.33\textwidth]{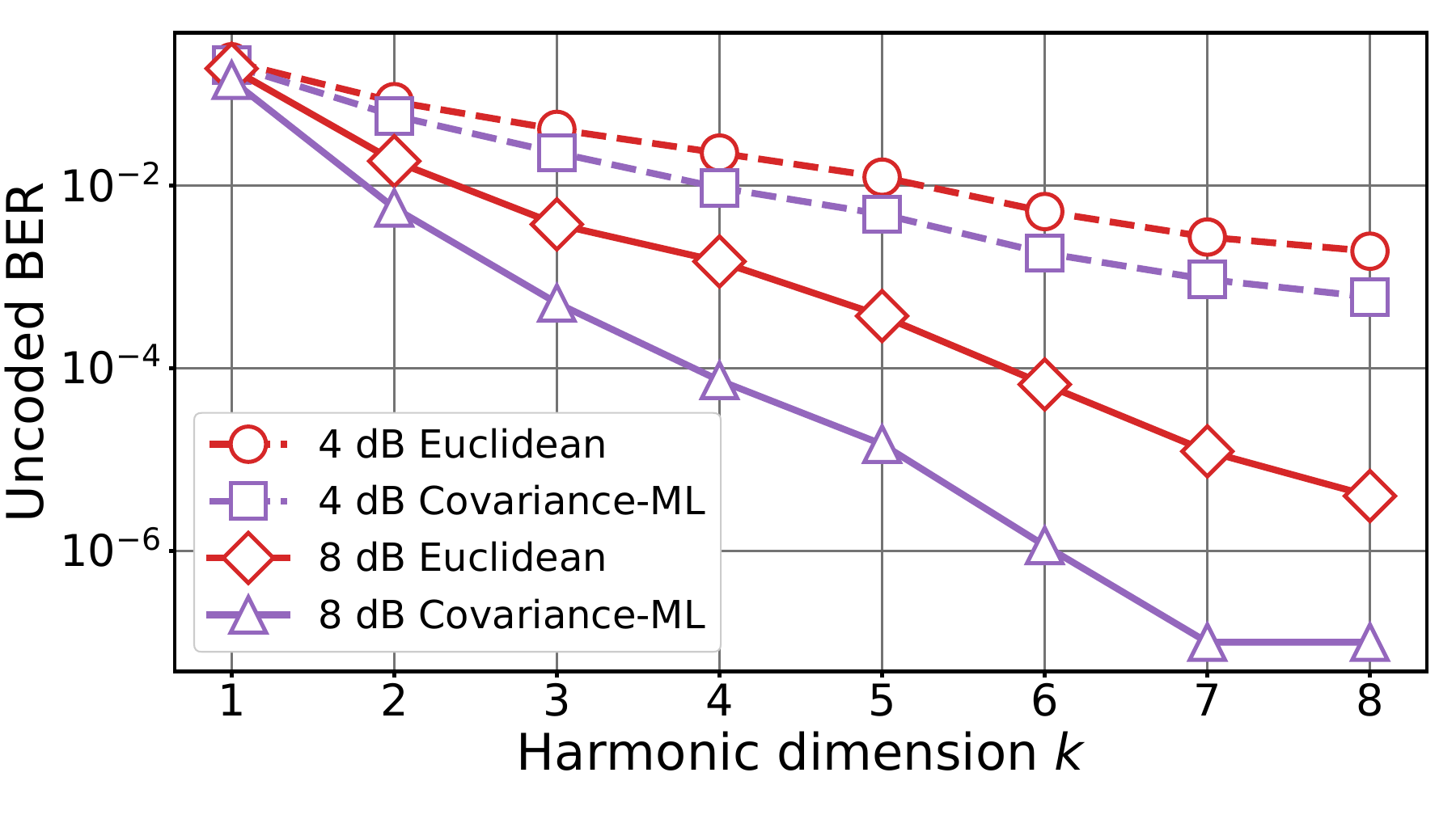}
\includegraphics[width=0.33\textwidth]{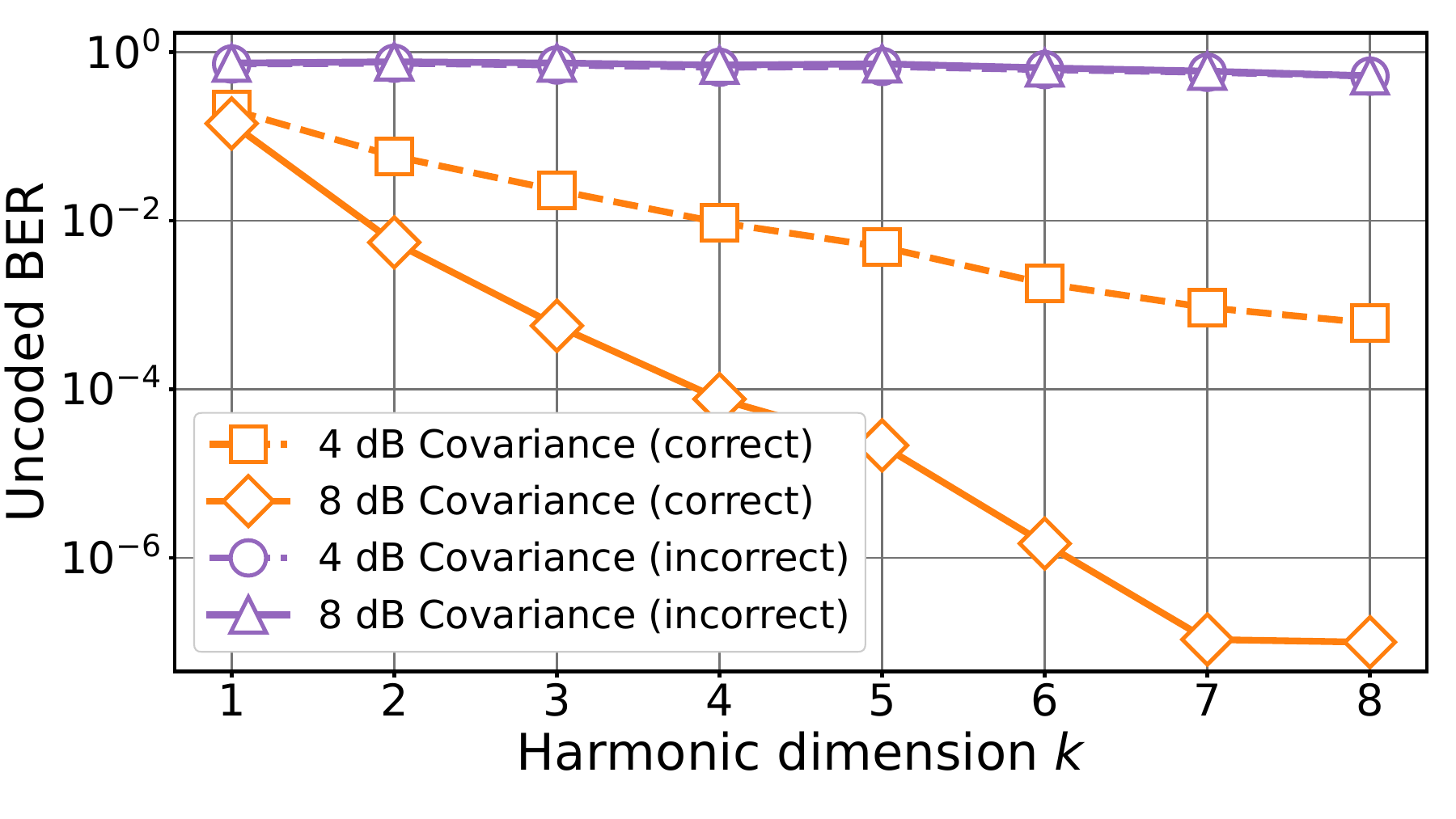}
\caption{\glsentryshort{ber} of \gls{qpsk} demodulation without \gls{ldpc}
versus harmonic dimension at $4$ and $8$ dB. Left: unkeyed, Euclidean vs.
covariance-aware demodulation. Right: keyed, covariance-aware demodulation
\textcolor{black}{with correct or incorrect codebook.} Curves are floored at $10^{-7}$, the
simulation resolution limit for $10^8$ transmitted bits per point. For $k>M$, these harmonic phases wrap modulo $2\pi$ and repeat cyclically.}
\label{fig:uncoded_qpsk_combined}
\end{figure*}
\begin{figure*}[!t] 
\centering
\includegraphics[width=0.33\textwidth]{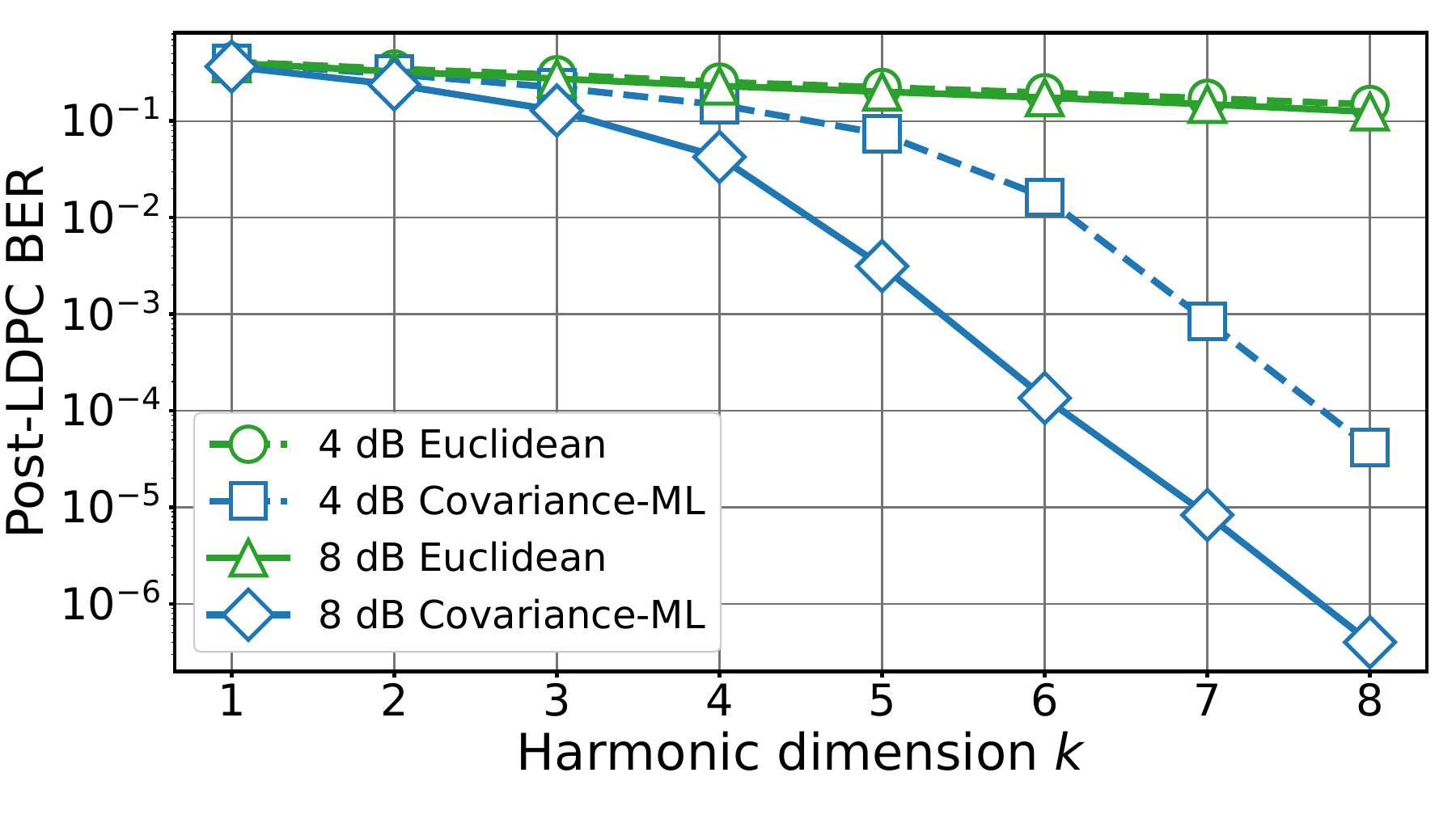}
\includegraphics[width=0.33\textwidth]{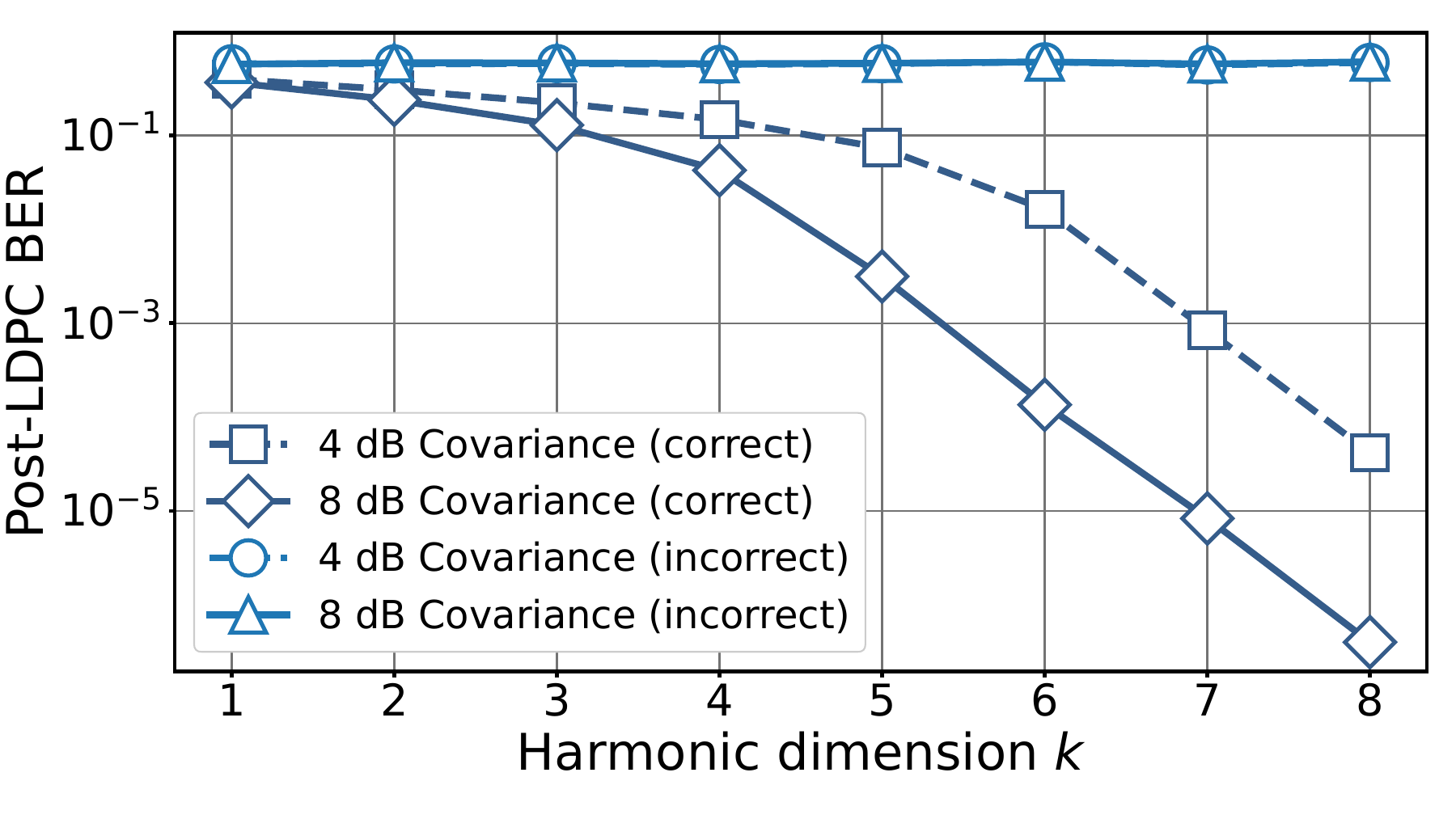}
\caption{\glsentryshort{ber} of 64-PSK with \gls{ldpc} decoding versus
harmonic dimension at $4$ and $8$ dB. Left: unkeyed, Euclidean vs.
covariance-aware demodulation. Right: keyed, covariance-aware demodulation
\textcolor{black}{with correct or incorrect codebook.}}
\label{fig:ldpc_combined}
\end{figure*}
\section{Numerical Results}
\subsection{\textcolor{black}{Communication BER Evaluation}}

\textcolor{black}{Here we evaluate communication \gls{ber}.} The sensing and control
channels use \gls{qpsk} for reliability and sensing performance, while the
payload channel uses higher-order modulation for throughput. We consider a
naive Eve with a Euclidean demapper and an architecturally informed Eve
(\gls{cu}) using a covariance-aware demapper but lacking the correct key. Independent Monte Carlo realizations are used for tangent
\gls{an} and phase keys, with default $\beta=0.5$, following the representative setting used
in~\cite{han2026matched}. \textcolor{black}{Throughout the numerical evaluation, \gls{snr} denotes the received \gls{snr} per active subcarrier.}

Figure.~\ref{fig:uncoded_qpsk_combined} shows an uncoded \gls{qpsk}
\textcolor{black}{harmonic-dimension sweep.} In general, increasing $k$ decreases \gls{ber} for both demodulators,
since the added harmonics enrich the signal's phase features and redundancy. The
covariance-aware demodulator outperforms the Euclidean baseline, and the gap
widens with increasing harmonic dimension, as the candidate-dependent noise
covariance becomes more pronounced and thus more exploitable at higher $k$.
Once the waveform is keyed, the \gls{ber} of a receiver lacking the
phase key
remains high even with the covariance-aware demodulator, while the
legitimate receiver, which possesses the key, sees no degradation in
performance. Figure~\ref{fig:ldpc_combined} compares the $1/2$ \gls{ldpc}-coded 64-PSK
decoding performance at $4$ and $8$ dB. The overall trends match the uncoded
case. Euclidean decoder performs worse compared to
the previous \gls{qpsk} scenario, which follows from the higher-order
constellation itself: with more candidate points packed into the same
unit-energy space, the minimum Euclidean distance between neighboring
symbols shrinks, making the Euclidean metric more sensitive to
noise and to the \gls{an} perturbation, independent of any keying effect.

\textcolor{black}{The evaluated receiver uses an unkeyed codebook without
estimating the phase key and exhibits high \gls{ber}.}
Higher $k$ reduces \gls{ber} but at the cost of spectral
efficiency. For sensing and control signals, which are typically short,
spectral efficiency is less relevant; control signals in particular require
\textcolor{black}{reliability}, motivating a high $k$ combined
with \gls{ldpc} coding for ultra-low \gls{ber}.

\subsection{Validation of waveform design and legitimate sensing}
\begin{figure}[!t]
\centering
\includegraphics[width=0.65\columnwidth]{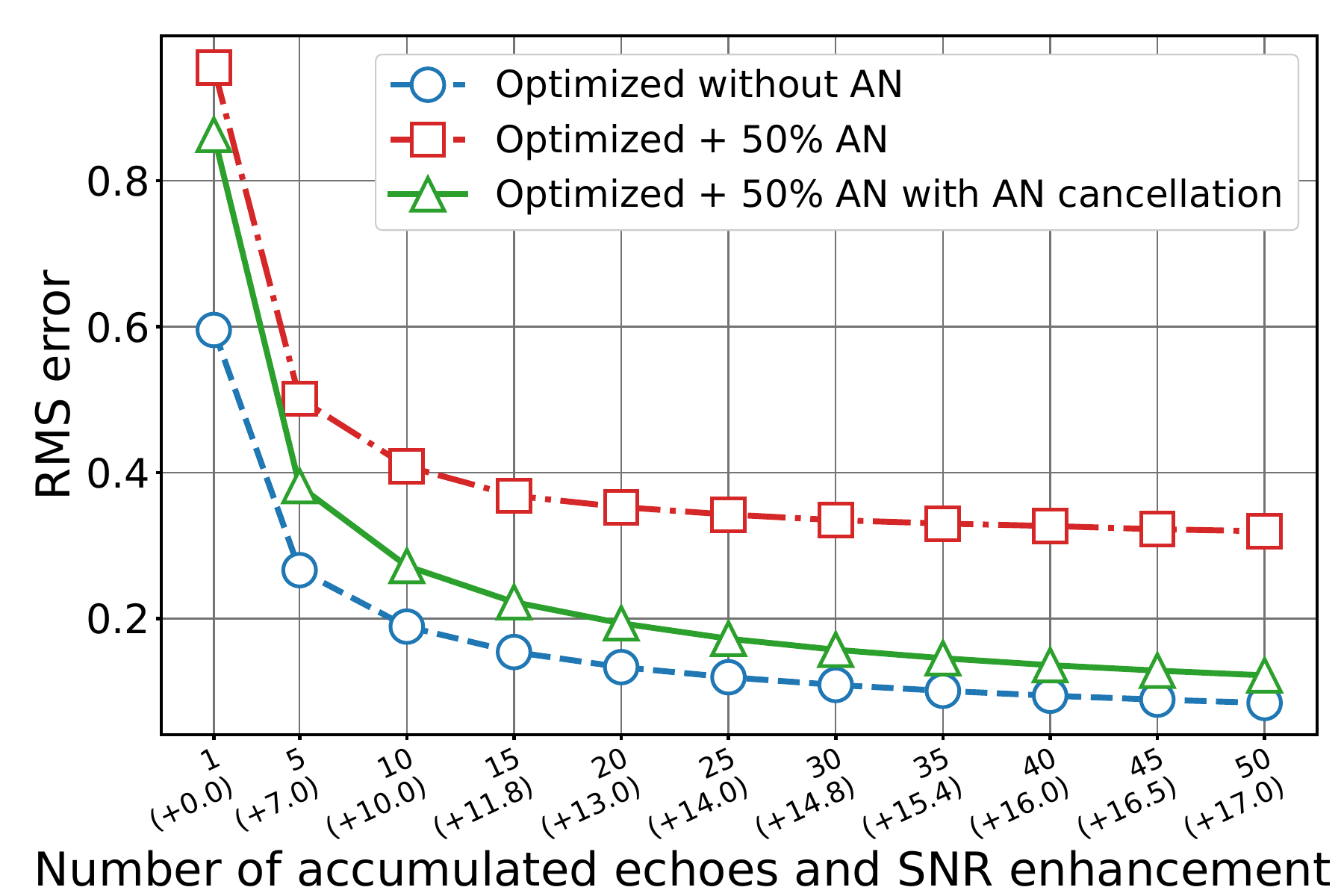}
\caption{Accumulated-echo RMS error for an optimized $k=5$ sensing waveform at
$6$ dB per-echo \glsentryshort{snr}. The receiver compensates
dominant known front-end and channel distortions. X-axis also reports the
\textcolor{black}{ideal \glsentryshort{snr} enhancement (dB) associated with $M$}}
\label{fig:echo_accum_rms}
\end{figure}

Next, we validate the sensing waveform design in a cooperating BS--BS case.
The waveform-only results in Table~\ref{tab:opt_vs_random_corr} show that the
optimized sequence achieves higher \gls{psr} than the random baseline, with
\gls{psr} around $20$~dB for most $k$. The dips at $k=4$ and $k=8$ are a
\textcolor{black}{structural effect of the QPSK alphabet: for $\phi_n = i\pi/2$, the harmonic
sum $\sum_{m=1}^{k} e^{jm\phi_n}$ has period $4/\gcd(i,4)$ in $m$ and
vanishes exactly when $k$ is a multiple of that period, leaving only
$\phi_n=0$ non-degenerate and sharply reducing waveform diversity.} For
$k>M$, these harmonic phases wrap modulo $2\pi$ and repeat cyclically.
\begin{table}[!t]
    \centering
    \caption{\glsentryshort{psr} (dB) of optimized and random
    \glsentryshort{qpsk} waveforms: $N=64$, \glsentryshort{snr} $=5$ dB,
    $\beta=0.5$, 2000 Monte Carlo runs.}
    \label{tab:opt_vs_random_corr}
    \scriptsize
    \setlength{\tabcolsep}{1.8pt}
    \renewcommand{\arraystretch}{1.05}
    \begin{tabular}{lcccccccc}
        \toprule[1.2pt]
        \rowcolor{gray!20}
        \textbf{$k$} & \textbf{1} & \textbf{2} & \textbf{3} & \textbf{4} & \textbf{5} & \textbf{6} & \textbf{7} & \textbf{8} \\
        \midrule
        \rowcolor{gray!8}
        \textbf{Opt. \glsentryshort{psr}} & 20.95 & 21.86 & 19.65 & 17.40 & 20.52 & 21.47 & 19.11 & 17.36 \\
        \rowcolor{gray!15}
        \textbf{Rand. \glsentryshort{psr}} & 16.29 & 13.90 & 13.51 & 7.67 & 12.29 & 10.80 & 12.16 & 8.67 \\
        \bottomrule[1.2pt]
    \end{tabular}
\end{table}
\textcolor{black}{We then quantify the reconstruction of sensing echoes.} Fig.~\ref{fig:echo_accum_rms} shows that the
reconstruction error curve converges toward the no-\gls{an} baseline,
indicating that coherent sensing can be systematically recovered
across accumulated echoes despite the added \gls{an}, residual timing error and channel distortion.

\subsection{validation of secure sensing}

Finally, we evaluate secure sensing performance, considering the threat
from Section~\ref{sec:thmodel} where Eve observes the transmitted signal
over a long period and attempts cross-burst coherent reference
reconstruction. We transmit the same nominal sequence multiple times with
independent \gls{an} realizations and phase keys.
\textcolor{black}{We use symbol-dependent keys refreshed for each burst.
Position-dependent and pseudorandom key-generation rules are studied in our
extension.}
\textcolor{black}{Eve cross-correlates the received signals to detect potential sensing signals,
\textcolor{black}{the detection rate is $P_C$.}} Two random-sequence baselines isolate the contributions of keying, \gls{an},
and waveform optimization: unprotected ($\textbf{Rand. No}$ in
Table~\ref{tab:keyed_burst_beta}), and keyed with \gls{an} $\beta=0.5$
($\textbf{Rand. A+K}$). As Table~\ref{tab:keyed_burst_beta} shows, the
unprotected optimized sequence ($\textbf{Opt. No}$) gives $P_C=1.0$ versus
$0.05$ for the unprotected random baseline. Keying alone ($\beta=0$) drops
$P_C$ to $0.386$; adding \gls{an} drops it further, falling below the
unprotected random baseline by $\beta\approx0.3$ and reaching $P_C=0$ from
$\beta\approx0.6$, matching the keyed-and-\gls{an} random baseline. To
isolate the role of key randomization itself, the $\textbf{Opt. A+SK}$
entry considers optimized bursts with the same phase key but
independent \gls{an} realizations ($\beta=0.5$); its high $P_C$ confirms that phase-key
updates, not \gls{an} alone, is essential for suppressing
$P_C$ for Eve. Notably, the keyed optimized \gls{psr} remains
much higher: \gls{psr} reflects correlation \emph{shape}, while $P_C$
depends on peak \emph{magnitude} relative to the noise floor. Independent
keys disrupt the constructive alignment behind the optimized peak,
collapsing its magnitude even as the shape (and \gls{psr}) stays favorable,
so high \gls{psr} cannot rescue detection once the mainlobe is suppressed.
The comparison of \textbf{Opt. $0.1\pi$} and \textbf{Opt. $0.25\pi$}
evaluates waveform-based detection when Eve has estimated the phase keys.
When the average phase-key error increases to $\sigma_\phi=0.25\pi$, Eve can
barely identify the sequence through waveform-based detection. With such low $P_C$, the sensing reference can be made
difficult for Eve to detect and reconstruct, or the system can
further randomize transmit timing and deliberately send keyed random
sequences with \gls{an} to confuse Eve.

\textcolor{black}{In Tab.~\ref{tab:key_mismatch_ser_k5}, Eve jointly evaluates phase-key
candidates and their sensing-symbol decisions from multiple
observations, then identifies the signal by comparing the recovered sequences.
Matching success generally decreases nonlinearly with \gls{ser}. We omit a
matching-error model and focus on low mismatch, neglecting this
loss in Eve's favor. For a uniform-grid search with resolution
$\Delta_\phi=\sigma_\phi\pi$, \textcolor{black}{each of the $kN$ independent phase-key
coefficients} has $Q\approx\lceil2/\sigma_\phi\rceil$ candidates, yielding
$Q^{kN}$ candidate key arrays. We assume that Eve scores each array by
demodulating all $N$ QPSK symbols. With four metrics per symbol, this grid
search costs
$C_{\rm Eve}=4NQ^{kN}=4N\lceil2/\sigma_\phi\rceil^{kN}$ metric evaluations
per burst. This count applies when Eve knows only the phase-key space, without
knowing how the keys are generated. If the keys follow the symbol-dependent
lookup-table rule of~\cite{han2026matched}, Eve can exploit this reuse to
perform the attack in~\cite{han2027reuse} without exhaustively enumerating
the key space.}
\begin{table}[!t]
    \centering
    \caption{Repeated-burst keyed \glsentryshort{qpsk} cross-correlation:
    $N=64$, $k=5$, \glsentryshort{snr} $=6$ dB,
    $P_{\mathrm{FA}}=0.05$.}
    \label{tab:keyed_burst_beta}
    \scriptsize
    \setlength{\tabcolsep}{4.1pt}
    \renewcommand{\arraystretch}{1}
    \begin{tabular*}{0.84\columnwidth}{@{\extracolsep{\fill}}C{0.12\columnwidth}ccc|C{0.12\columnwidth}ccc}
        \toprule[1.2pt]
        \rowcolor{gray!20}
        \textbf{$\beta$} & \textbf{Peak} & \textbf{\glsentryshort{psr}} & \textbf{$P_{C}$}
        & \textbf{$\beta$} & \textbf{Peak} & \textbf{\glsentryshort{psr}} & \textbf{$P_{C}$} \\
        \midrule
        \rowcolor{gray!5}
        0.0 & 0.262 & 5.620 & 0.386 & 0.5 & 0.143 & 2.350 & 0.002 \\
        \rowcolor{gray!5}
        0.1 & 0.235 & 5.280 & 0.172 & 0.6 & 0.124 & 1.480 & 0.000 \\
        \rowcolor{gray!5}
        0.2 & 0.211 & 4.890 & 0.074 & 0.7 & 0.116 & 1.000 & 0.000 \\
        \rowcolor{gray!5}
        0.3 & 0.187 & 4.160 & 0.016 & 0.8 & 0.112 & 0.810 & 0.000 \\
        \rowcolor{gray!5}
        0.4 & 0.160 & 3.070 & 0.002 & 0.9 & 0.112 & 0.690 & 0.000 \\
        \midrule
        \rowcolor{gray!15}
        $\textbf{Rand. No}$& 0.202 & 0.780 & 0.050 & $\textbf{Rand. A+K}$& 0.105 & 0.626 & 0.000 \\
        \rowcolor{gray!15}
        \midrule
        \rowcolor{gray!8}
        $\textbf{Opt. No}$& 0.725 & 6.880 & 1.000 & $\textbf{Opt. A+SK}$& 0.369 & 6.976 & 0.925\\ \midrule
        $\textbf{Opt 0.1$\pi$}$& 0.331 & 6.942 & 0.886 & $\textbf{Opt 0.25$\pi$}$& 0.203& 2.762 & 0.051 \\
        \bottomrule[1.2pt]
    \end{tabular*}
\end{table}

\begin{figure}[!t]
\centering
\includegraphics[width=0.75\columnwidth]{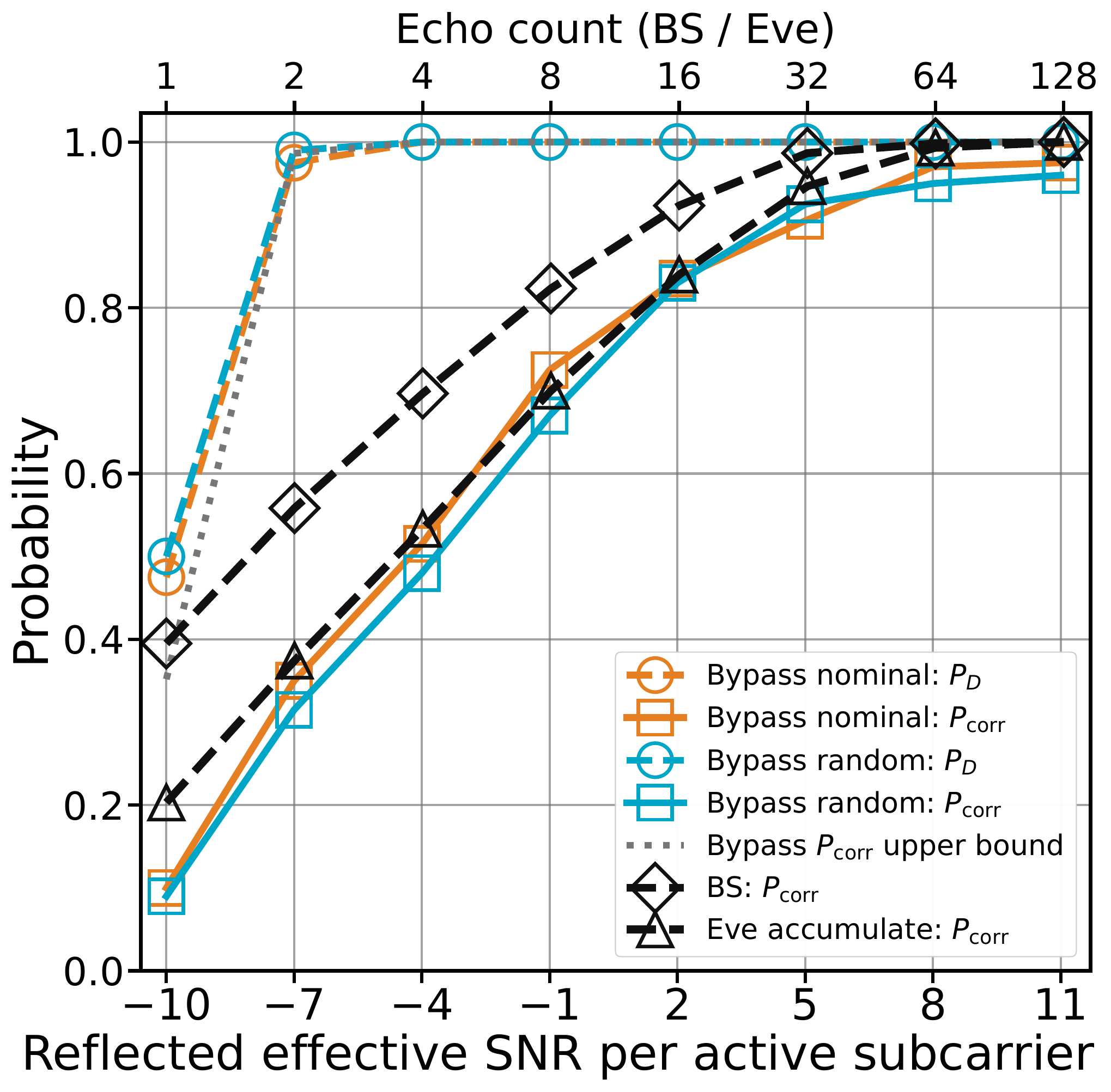}
\caption{Bypass detection and exact-bin locking: $k=5$, $\beta=0.5$.
Black dashed curves compare ideal BS and Eve with imperfect key recovery
locking with $-10$ dB useful \glsentryshort{snr} per echo; echo counts
appear above. The lower axis denotes received \glsentryshort{snr} for
bypass and effective useful \glsentryshort{snr} for BS/Eve.}
\label{fig:pd_k2_k6_compare}
\end{figure}

\textcolor{black}{Next, Fig.~\ref{fig:pd_k2_k6_compare} considers a wireless-channel
vulnerability: Eve can use the direct-path observation as a reference to detect
the reflected signal, bypassing the need to know or decode the sensing waveform.}
For tractability, we assume Eve always has \gls{los} to both the
transmitter and target, with the direct path $4$--$8$~dB stronger than
the reflected path. We evaluate the sensing reference against a
random-sequence baseline, both keyed with \gls{an}, where direct- and
reflected-path observations from the same burst share the same key and
\gls{an} realization. In Fig.~\ref{fig:pd_k2_k6_compare},
both $P_{\mathrm{corr}}$ and $P_D$ are similar for the random and the
optimized sequences. \textcolor{black}{This follows because Algorithm~1 optimizes the
mainlobe/sidelobe trade-off for the clean, unkeyed, \gls{an}-free case
through a sensitive phase alignment across chips; keying and \gls{an}
disrupt this alignment severely, bringing performance down to the level
of a random sequence of the same length.}

\textcolor{black}{However, Eve's bypass-attack performance can be limited by
controlling transmit power.}
%Non-coherent detection is therefore unnecessary for the legitimate receiver:
%once it demodulates the reference, it locates the sensing sequence
%and integrate power across bursts, achieving the sensing performance that Eve, cannot replicate.
\textcolor{black}{For \textbf{Eve without phase-key reconstruction}, independent
key refreshing disrupts a fixed cross-burst waveform reference, as illustrated
in Table~II. Without inter-burst phase alignment, Eve instead uses noncoherent
accumulation, which discards phase information and generally requires more
observations to achieve the detection performance of coherent accumulation.}
For \textbf{Eve with phase-key reconstruction}, with
\textcolor{black}{$\sigma_\phi=0.1\pi$} and assuming sequence-mismatch error is neglected,
Eve's $P_{\mathrm{corr}}$ is generally lower than that of the legitimate
receiver. This stems from two
compounding effects of the phase-key mismatch: the residual phase error
directly degrades correlation, and it simultaneously worsens \gls{an}
cancellation, further corrupting the observation. Both effects are structural.
\textcolor{black}{At $\sigma_\phi=0.1\pi$, exhaustive uniform-grid search over
an unrestricted $N\times k$ phase mask requires $20^{320}$ candidate arrays
and approximately $5.5\times10^{418}$ metric evaluations per burst.
Independent key refreshing prevents reuse of the recovered mask across bursts.}

\begin{table}[!t]
    \centering
    \caption{\glsentryshort{ser} and blind-search cost under phase-key mismatch:
    $N=64$, $k=5$, \glsentryshort{snr} $=1$ dB, $\beta=0.5$.
    Zero mismatch denotes the legitimate exact-key reference.}
    \label{tab:key_mismatch_ser_k5}
    \scriptsize
    \setlength{\tabcolsep}{0.8pt}
    \renewcommand{\arraystretch}{1.05}
    \begin{tabular}{@{}>{\centering\arraybackslash}p{0.11\columnwidth}*{6}{>{\centering\arraybackslash}p{0.135\columnwidth}}@{}}
        \toprule
        \rowcolor{gray!20}
        \textbf{$\sigma_\phi$ ($\pi$)} &
        \textbf{0} & \textbf{0.05} & \textbf{0.10} & \textbf{0.15} &
        \textbf{0.20} & \textbf{0.25} \\
        \midrule
        \rowcolor{gray!8}
        \textbf{SER} &
        0.051 & 0.060 & 0.074 & 0.112 & 0.152 & 0.238 \\
        \rowcolor{gray!15}
        \textbf{$C_{\rm Eve}$} &
        -- & {\tiny \mbox{$1.2\times10^{515}$}} & {\tiny \mbox{$5.5\times10^{418}$}} &
        {\tiny \mbox{$1.5\times10^{369}$}} & {\tiny \mbox{$2.6\times10^{322}$}} &
        {\tiny \mbox{$2.5\times10^{291}$}} \\
        \bottomrule
    \end{tabular}
    \vspace{2pt}

    \begin{tabular}{@{}>{\centering\arraybackslash}p{0.11\columnwidth}*{6}{>{\centering\arraybackslash}p{0.135\columnwidth}}@{}}
        \toprule
        \rowcolor{gray!20}
        \textbf{$\sigma_\phi$ ($\pi$)} &
        \textbf{0.30} & \textbf{0.35} & \textbf{0.40} &
        \textbf{0.45} & \textbf{0.50} & \textbf{--} \\
        \midrule
        \rowcolor{gray!8}
        \textbf{{SER}} &
        0.289 & 0.377 & 0.529 & 0.527 & 0.581 & -- \\
        \rowcolor{gray!15}
        \textbf{$C_{\rm Eve}$} &
        {\tiny \mbox{$6.9\times10^{272}$}} & {\tiny \mbox{$2.6\times10^{251}$}} &
        {\tiny \mbox{$1.2\times10^{226}$}} & {\tiny \mbox{$1.2\times10^{226}$}} &
        {\tiny \mbox{$1.2\times10^{195}$}} & -- \\
        \bottomrule
    \end{tabular}
\end{table}

\section{Discussion and Conclusion}

This paper studied waveform randomization for protecting sensing references
in \gls{isac} systems using Fourier-curve modulation, phase keys, and
tangent \gls{an}. The results show that Eve's sensing capability can be
strongly suppressed even when she exploits both direct- and reflected-path
observations under favorable geometry. We further considered phase-key
\textcolor{black}{reconstruction by Eve and showed that residual key mismatch and imperfect
\gls{an} removal still limit coherent sensing.}

Future work will investigate joint optimization of the phase key,
tangent-\gls{an} power, and harmonic dimension $k$ under explicit sensing
accuracy and communication reliability constraints, and will validate the
design in multi-target and multi-Eve scenarios closer to practical B5G/6G
deployments.
\bibliographystyle{IEEEtran}
\bibliography{references}

\end{document}